\documentclass[english,twocolumn,aps,prd,longbibliography,crop,tikz,superscriptaddress]{revtex4-1}
\usepackage[T1]{fontenc}
\usepackage{graphicx}
\usepackage{babel}
\usepackage{braket}
\usepackage{qcircuit}
\usepackage[dvipsnames]{xcolor}
\usepackage{amsmath}

\usepackage{qcircuit}
\usepackage[colorlinks,citecolor=red,urlcolor=blue,bookmarks=false,hypertexnames=true]{hyperref} 

\usepackage{fancyhdr}

\usepackage{fancyhdr}
\begin{document}

\title{Simulating spectroscopy experiments with a superconducting quantum computer}

\author{John P. T. Stenger}
\affiliation{NRC Postdoctoral Associate, U.S. Naval Research Laboratory, Washington, DC 20375, United States}
\author{Gilad Ben-Shach}
\affiliation{IBM Quantum, IBM Canada, Markham, Ontario, L6G 1C7, Canada}
\author{David Pekker}
\affiliation{Department of Physics and Astronomy, University of Pittsburgh, Pittsburgh, PA 15260, USA}
\author{Nicholas T. Bronn}
\affiliation{IBM Quantum, IBM T.J. Watson Research Center, Yorktown Heights, New York 10598, USA}

\begin{abstract}

We present a novel method for solving eigenvalue problems on a quantum computer based on spectroscopy.  The method works by coupling a ``probe'' qubit to a set of system simulation qubits and then time evolving both the probe and the system under Hamiltonian dynamics.  In this way, we simulate spectroscopy on a quantum computer.  We test our method on the IBM quantum hardware for a simple single spin model and an interacting Kitaev chain model.  For the Kitaev chain, we trace out the pseudo-topological phase boundary for a two-site model. 

\end{abstract}

\maketitle

\thispagestyle{fancy}

The simulation of quantum systems with a quantum computer was originally proposed by Feynman~\cite{Feynman1982} which sparked the nascent field. While the dynamics of quantum systems are fully determined (e.g. by the Schr\"odinger equation), the dimension of the Hilbert space that describes their evolution scales exponentially in the number of degrees of freedom which makes classical calculations impossible for a large enough system size.  The recent availability of quantum computers on the cloud allows practitioners to begin to map problems of quantum simulation and achieve proof-of-principle results~\cite{Stenger2021, Xiao2021, Rancic2021, Sung2022, Harle2022, Mi2022}, even starting to surpass the accuracy of state-of-the-art classical approximations~\cite{Kim2021}. While quantum phase estimation (QPE)~\cite{Kitaev1995} offers an exponential speed-up in determining Hamiltonian spectra, it requires a depth of circuit unfeasible for near-term noisy quantum hardware. Short-depth hybrid algorithms such as the variational quantum eigensolver (VQE) are amenable to current hardware at the expense of many iterations over measurements and optimization cycles~\cite{Peruzzo2014, Kandala2017, McClean2016, OMalley2016, Wang2019, Gonthier2020, McArdle2020}. Furthermore, VQE requires a problem-specific ansatz which is not generalizable to all systems. Here we propose a spectroscopic eigensolver method which makes an efficient use of available noisy near-term quantum resources. Our method is both amenable to noisy quantum hardware and is generally applicable to determining Hamiltonian spectra.


\begin{figure}[b!]
\begin{center}
\includegraphics[width=\columnwidth]{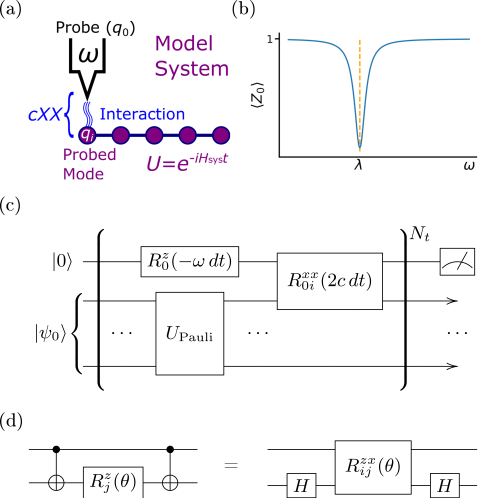}
\end{center}
\vspace{-2mm}
\caption{Quantum circuits for simulating the time evolution $U_{\rm Pauli} = e^{-iH_{\rm Pauli} t}$: (a) a cartoon analogy of the spectroscopy eigensolver method to scanning tunneling spectroscopy, where a tip (qubit) at potential (energy) $\omega$ interacts with a single mode of a model system via electron tunneling (transverse $cXX$ interaction), revealing (b) a response as a function of energy which determines an eigenvalue $\lambda$ as a dip in probe expectation $\langle Z_0 \rangle$. (c) The spectroscopic eigensolver algorithm for a probe qubit interacting with a single qubit of the simulation register (for $N_t$ Trotter steps $U_{\rm res}$, first-order shown for clarity) followed by measurement of the probe qubit, and (d) a circuit equivalence that allows the efficient implementation of scaled pulses with the native $R^{zx}(\theta)$ interaction generated by echoed cross resonance~\cite{Sheldon2016}.
\label{spec-eigen-circ}}
\end{figure}


We present a quantum eigensolver method which is akin to the experimental tool of tunneling spectroscopy. Our method involves a ``probe qubit'' and a ``simulation'' register that undergo Hamiltonian evolution for a time $t$.  A cartoon of our method is depicted in Fig.~\ref{spec-eigen-circ}a.  Measurement of the ``probe'' qubit $q_0$ reveals the spectrum of the evolved Hamiltonian via its response when the probe energy is on-resonance with an energy transition in the system, schematically shown as a dip in Fig.~\ref{spec-eigen-circ}b.  While similar to other time-evolution algorithms~\cite{Kitaev1995, Somma2002}, this technique does not require controlled-coherent evolution of the register by auxiliary qubits. Instead, the probe qubit evolving according to an energy (via $\omega$)  interacts tranversely with a single qubit of the simulation register. This process is repeated over a number of Trotter steps as shown in Fig.~\ref{spec-eigen-circ}c.  We have the freedom to select the specific qubit being probed. It is only necessary that the wavefunction of the simulation register has strong weight on that qubit. This method is conducive for execution on near-term noisy quantum hardware because it only requires a local interaction between qubits instead of the controlled-evolution of an entire simulation register, substantially reducing circuit depth because no SWAP operations are necessary between probe qubit and simulation register and naturally maps to the planar architecture of current superconducting quantum hardware (see Supplement~I).  Furthermore, only measurement of the probe qubit is required, reducing readout error.  The underlying trade-off as compared to other algorithms is the number of circuits that must be executed: one for each value of $\omega$, a sweep of which is required for accurate fitting and hence determination of the eigenvalues (see Supplement II). However, recent improvements as determined by speed benchmarking show this is not a bottleneck for superconducting quantum hardware~\cite{Wack2021}. 

Explicitly, the spectroscopic eigensolver technique finds the energy differences of a system by coupling a probe qubit to a Pauli Hamiltonian $H_{\rm Pauli}$ suitably encoded to enforce the commutation relations of the  system Hamiltonian $H_{\rm sys}$ it represents. The full Hamiltonian, which we will refer to as the ``resonance'' Hamiltonian because of its response when the probe and system are on-resonance, is
\begin{equation}
    H_{\rm res} = -\frac{1}{2}\omega Z_0 + c X_0 X_i + H_{\rm Pauli},
    \label{eq:res-eq}
\end{equation}
where the probe qubit $q_0$ with energy $\omega$ is at index 0, $c$ is the coupling parameter between the probe and probed qubit $q_i$ of the simulated system, and the tensor products with identity matrices are omitted for succinctness.  The first-order Suzuki-Trotter decomposition of the time evolution unitary is 
\begin{equation}
    U_{\rm res} (\omega, dt) = e^{-iH_{\rm res} dt} \approx R^{xx}_{0i}\left(2c\,dt\right) R^z_0\left(-\omega\,dt\right) U_{\rm Pauli},
\end{equation}
where $R^z_j(\phi)$ is a $Z$-rotation on site $j$, $R^{xx}_{jk}(\theta)$ is an $XX$-rotation on sites $j$ and $k$ and $U_{\rm Pauli} = e^{-i H_{\rm Pauli} dt}$ is the Trotter step of time evolution $dt = t/N_t$ for the system. 

We start in a state $\ket{0,\psi_0}$ with the probe qubit in the ground and the system $\ket{\psi_0}$ is in an arbitrary state.  For simplicity we will take $|\psi_0\rangle$ to be the $n$-qubit ground state. After $N_t$ applications of $U_{\rm res}$, we measure the probe qubit in the $Z$-basis, $\braket{Z_0}(\omega) = \bra{0,\psi_0} (U_{\rm res}^\dagger)^{Nt} Z_0 U_{\rm res}^{Nt} \ket{0,\psi_0}$.   When $c$ is small compared to the energy scales in $H_{\rm Pauli}$ and when $\omega$ is on resonance with an energy transition of $H_{\rm Pauli}$, the probability of the probe qubit flipping will peak, resulting in a dip in $\braket{Z_0}(\omega)$. 

\begin{figure}[t]
\begin{center}
\includegraphics[width=\columnwidth]{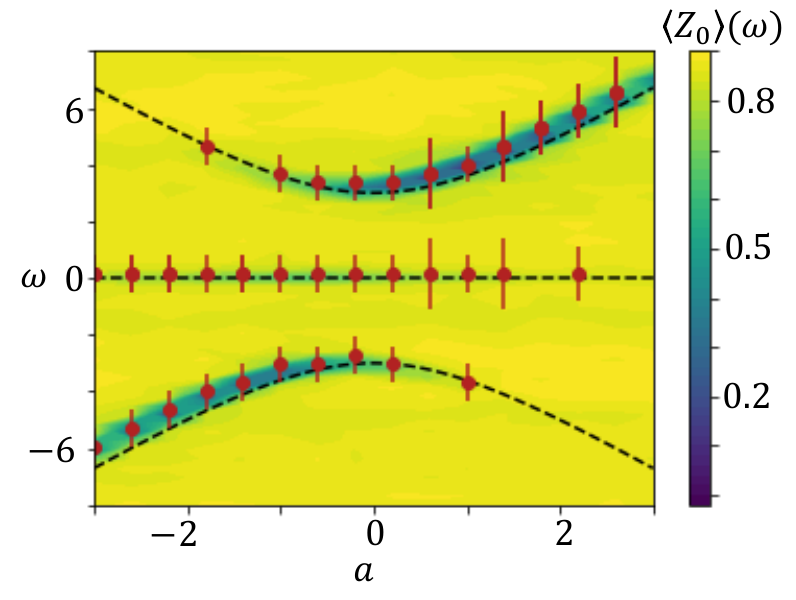}
\end{center}
\vspace{-2mm}

\caption{Scans of the probe qubit expectation value as a function of $\omega$ for many values of $a$.  The minimums in $\braket{Z_0}(\omega)$ for each value of $a$ are plotted as red circles.  The error bars represent the full-width-at-half-minimum for the $\braket{Z_0}(\omega)$ dips.  The black dashed lines are the expected value of the energy transitions as obtained from exact diagonalization. The parameter values are $c = 0.1$, $t = 10$, $dt = 0.3\bar{3}$.}
\label{FLZ}
\vspace{3mm}
\end{figure}


The following experiments were conducted within the Qiskit~\cite{Qiskit} framework with jobs sent to the cloud-based {\it ibm\_lagos}, a 7-qubit superconducting IBM Quantum backend. First, the system Hamiltonian is mapped to a qubit (Pauli) Hamiltonian by a suitable encoding. The Pauli Hamiltonian is then mapped to unitary time evolution via the second-order Suzuki-Trotter transformation. The resulting circuit consists of two-qubit rotations on the order of $c\,dt$, $\omega\,dt$, etc., where the digital synthesis of each is locally equivalent to a $Z$-rotation sandwiched by two CNOTs, which is more efficiently implemented by scaled gates that rotate smaller angles in the two-qubit Hilbert space natively (the equivalence of which is shown in Fig.~\ref{spec-eigen-circ}d) instead of the (net) two-qubit $\pi$ rotation in the former case~\cite{Stenger2021}. These scaled pulses are implemented as calibrations for $R^{zx}(\theta)$ gates following an analysis of the circuit using a novel template optimization technique~\cite{Iten2019}. Each spectroscopic experiment is then built as an array of circuits, one for each value of $\omega$. For each circuit, a result consisting of the measurement strings of 8192 shots is returned and analyzed. This allows us to find the energy transitions by locating dips in $\braket{Z_0}(\omega)$.  Further details are provided in the Supplementary Material and full data and source code can be found at \url{www.github.com/qiskit-research/qiskit-research}.

As a first test case, we take the system to be a single spin under two opposing magnetic fields,  
\begin{equation}
    H_{\text{Pauli}} = H_{\text{sys}} = a Z_1 + b Y_1.
\end{equation}
We treat $b$ as fixed and vary $a$ so that the energy levels form an avoided crossing.  Both $a$ and $b$ are treated as unitless algorithmic quantities.  In Fig.~\ref{FLZ} we show scans of the orientation of the probe qubit $\braket{Z_0}(\omega)$ for several values of $a$. Notice that there is a dip at $\omega = 0$.  This corresponds to the transition of an energy level to itself, see the Supplement~III for details.  

We find excellent agreement between the exact energy transitions and the local minimum in $\braket{Z_0}(\omega)$.  The Root-Mean-Square (RMS) of the variance of all three transitions together is $\text{RMS} = \sqrt{\frac{1}{N_d}\sum_{i=0}^{N_d}(\omega^*_i - \Delta E_i)^2} \approx 0.083$ where $N_d$ is the number of data points, $w^*_i$ is a minimum and $\Delta E_i$ is the corresponding energy transition.  The RMS variance is about an order of magnitude less than the average uncertainty calculated from the Full-Width-at-Half-Minimum (FWHM), which includes both hardware and Trotter error, $ \text{FWHM} = \frac{1}{N_d}\sum_{i=0}^{N_d}(\omega^L_i - \omega^R_i) \approx 0.811 $
where $\omega^L_i$ is the value of $\omega$ where $\braket{Z_0}(\omega)$ is half way between its minimum and maximum value to the left of $\omega^*_i$ and $\omega^R_i$ is the same but to the right of $\omega^*_i$.  This simple test system stands to demonstrate how accurate our algorithm can be when the decoherence time of the qubits ($\sim 100-150$~$\mu$s) allows for a long evolution time $t \gg dt$, where $t$ corresponds to $\sim 15~\mu$s of hardware time for this example.  With this in mind, we will push our algorithm to the limits of the quantum hardware by testing a more complicated system.

A system of particular interest in tunneling spectroscopy is the $n$-site Kitaev chain which hosts Majorana Zero Modes (MZMs) in the topological regime.  The Kitaev chain is a particularly good test case for the spectroscopic eigensolver technique since the MZMs exist at the ends of the chain and are, therefore, easy to probe. Distinguishing the topological phase from the trivial superconducting phase in such nanowires has been a recent theoretical interest largely motivated by the fact that Majorana zero modes promise application in topologically protected quantum computation  \cite{Kitaev2003,Nayak2008,Alicea2011,Bonderson2011,Stenger2019,Ben-Shach2015}.  The topological regime is characterized by a change in parity of the ground state.  While the topological regime is only truly present in a $n\rightarrow \infty$ limit, a parity flip occurs even in a two site model. 

The model Hamiltonian of the interacting Kitaev chain with Coulomb interactions is 

\begin{equation}
    \begin{split}
    H_{\rm sys} = &\mu \sum_{i=1}^{L}c^{\dagger}_i c_i + g\sum_{i=1}^{L-1} (c^{\dagger}_ic_{i+1} + c^{\dagger}_{i+1}c_i) 
    \\
    + &\Delta \sum_{i=1}^{L-1} (c^{\dagger}_ic^{\dagger}_{i+1} + c_{i+1}c_i) + V \sum_{i=1}^{L-1} c^{\dagger}_ic_ic^{\dagger}_{i+1}c_{i+1}
    \end{split}
    \label{eq:kitaev-ham}
\end{equation}
where $L$ is the number of sites, $c^{\dagger}_i$ ($c_i$) is the creation (annihilation) operator on site $i$, $\mu$ is the chemical potential, $g$ is the hopping rate, $\Delta$ controls the superconductivity pairing strength, and $V$ is the interaction strength.  The inclusion of the interaction term makes the system difficult to study classically but poses no additional fundamental challenges for quantum processors~\cite{Childs2018, Tacchino2020}.

Using the Jordan-Wigner encoding~\cite{Somma2002}, we can express Eq.~\ref{eq:kitaev-ham} as
\begin{equation}
    \begin{split}
    H_{\rm Pauli} = &x \sum_{i=1}^{n-1} X_i X_{i+1} + y \sum_{i=1}^{n-1} Y_iY_{i+1} + z \sum_{i=1}^{n-1} Z_i Z_{i+1}
    \\
    & - m\sum_{i=1}^{n}Z_i - \bar{m}\sum_{i=2}^{n-1}Z_i 
    \end{split}
    \label{eq:pauli-ham}
\end{equation}
where $X_i,Y_i,Z_i$ are Pauli matrices acting on qubit $i$, $n=L$ is the number of qubits used to represent the system, and $2x = g+\Delta$, $2y = g-\Delta$, $4z = V$, $4m = 2\mu+V$, and $4\bar{m} = V$.  These parameters are considered to be unitless algorithmic quantities, however, the values we will use roughly correspond to those of real systems if they are taken to be in the meV range~\cite{Chen2017}.

In the limit $n\rightarrow \infty$, this system has a topological phase transition in parameter space.  We will restrict ourselves to the two-site $n=2$ model in the main text.  See Supplement~VI for an explanation of how to relate the two site model and the $n\rightarrow\infty$ limit and see Supplement~VIII for experimental results for the $n=3$ model.  In the $n=2$ case the gap between the ground and first excited state closes along the surface
\begin{equation}
    m = \sqrt{z^2 + z(x+y)+ xy}
    \label{zerocross}
\end{equation}
see Supplement~VII for details.  This surface can be thought of as a remnant of the topological phase boundary.  We now explore the use of the novel spectroscopic eigensolver technique to analyze this surface.

Because the circuit depth is long in this case, it is important that we optimize the parameters.  We optimized the time $t$, time step $dt$ and coupling $c$ (see Supplement~IV).  In Fig.~\ref{Fnew}a we show a sweep of the probe qubit's energy $\omega$ for three values of $c$.  The energy difference between the ground and first excited state is determined by the $\omega$ at which $\langle Z_0 \rangle (\omega)$ is minimum. When $c$ is too low, the probe qubit will not respond to the system but if c is too high it will perturb the system.  Note the state transition from ground to first excited (denoted by the notation [0,1]) is visible in Fig.~\ref{Fnew}a but the reverse transition is not. This is due to the initial state of the system qubits, set to $\ket{00}$ for simplicity, which has no overlap with the odd parity excited state and so only the forward transition is possible.  Considering a single parity, in this way, allows maximum contrast in our signal.   Absent also is the $\omega = 0$ dip which is true for all fermionic Hamiltonians  (see Supplement~III), making them particularly amenable to our method.  At the edges of
Fig.~\ref{Fnew}a we see the onset of higher energy transitions.  


\begin{figure}[t]
\begin{center}
\includegraphics[width=\columnwidth]{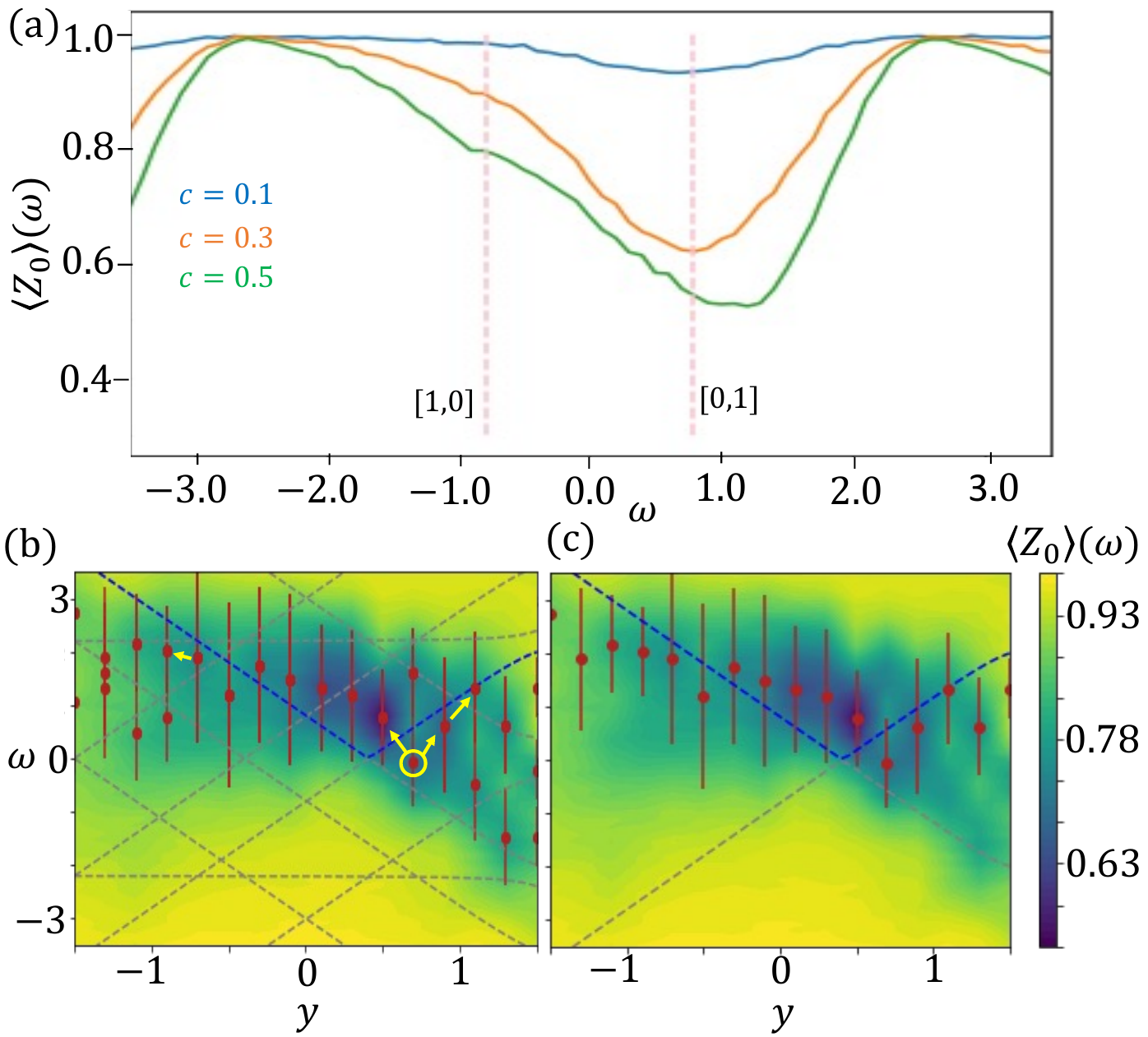}
\end{center}
\vspace{-2mm}

\caption{Scans of the probe qubit expectation value as a function of $\omega$.  (a) sweeps of the probe qubit frequency for different values of $c$.  For these sweeps, $m=1.0$, $x=1.5$, $y=0.4$, $z=0.2$, $dt = 0.7$ and $t=5$.  The vertical lines are the expected transitions from the classical solution.  For more $c$-sweeps see the Supplement~IV.  (b-c) sweeps of the probe qubit frequency for different values of $y$.  For these sweeps, $c = 0.3$, $m=0.1$, $x=1.5$, $z=0.4$, $dt = 0.7$ and $t=5$.  The minimums in $\braket{Z_0}(\omega)$ for each value of $y$ are plotted as red circles.  The error bars represent the half-width-at-half-minimum for the $\braket{Z_0}(\omega)$ dips.  The blue dashed line represents the [0,1] transition while other transitions are in grey. In panel (b) we show all transitions and all local minimums of $\braket{Z_0}(\omega)$.  In panel (c) we show only the [0,1] and [1,0] transition and we select the minimums by tracking as depicted by the yellow arrow in (b).}
\label{Fnew}
\vspace{3mm}
\end{figure}

Scans of  $\braket{Z_0}(\omega)$ for several values of $y$ are shown in Fig.~\ref{Fnew}b-c. In panel~\ref{Fnew}b the location of all local minima are displayed.  In \ref{Fnew}c we isolate the [0,1] transition by locating the minimum which is closest to $\omega = 0$ (circled in yellow in panel-b) and then track the dip by going to the neighboring values of $y$ and finding the minimum which is closest to the last.  The average FWHM $\approx 2.35$ is larger than the separation of some energy transitions.  This can cause dips from two different energy transitions to merge into one, shifting the location of the dip.  Even with this concern, we get fairly accurate results.  The RMS variance $\approx 1.72$ for the [0,1] transition which is well within the width of the peak. 


\begin{figure}[t]
\vspace{2mm}
\begin{center}
\includegraphics[width=\columnwidth]{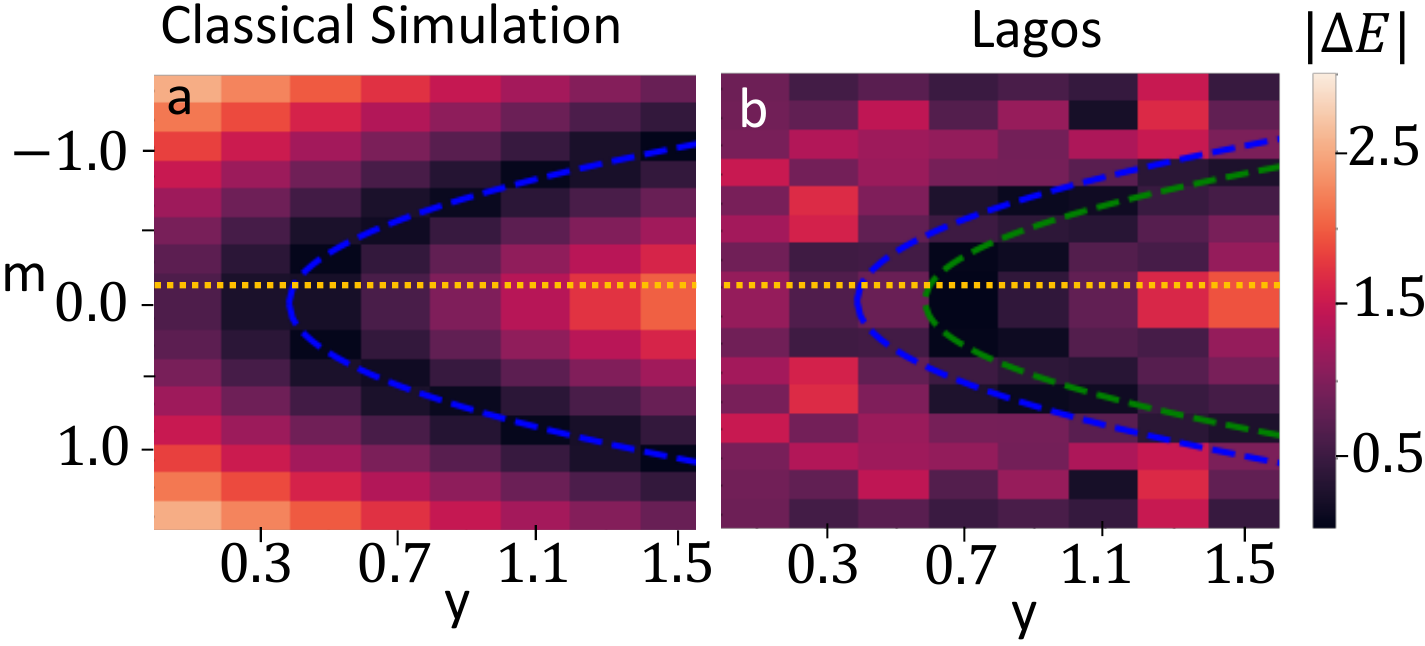}
\end{center}
\vspace{-2mm}

\caption{Energy difference between the ground and first excited state as a function of $m$ and $y$.  (a) Simulation on a classical computer with no error.   (b) results from the {\it ibm\_lagos} backend.  In both plots, $c=0.3$, $x=1.5$, $z=0.4$, $dt = 0.7$, and $t=5.0$. In both panels, the orange line shows the location of the data in Fig.~\ref{Fnew}b,c and the blue curve is the expected zero-crossing from Eq.~\ref{zerocross} with the above parameters.  The green curve in (b) is the zero-crossing again using Eq.~\ref{zerocross} with $z$ as the free parameter. We find the best-fit $z$ is shifted by $\Delta z = 0.19$.}
\label{phase-boundary}
\vspace{-3mm}
\end{figure}

Taking several scans of $\omega$ for different values of $m$ and $y$, we are able to trace out the pseudo phase boundary.  Fig.~\ref{phase-boundary} shows the absolute value of the gap for (a) a classical simulation without noise and (b) the results from the quantum device.  The curve of points at which the gap goes to zero is the phase boundary.  Notice that in the results from the quantum device the boundary is pushed to higher values of $y$ than in the perfect simulation.  The green line is a best fit of the boundary to Eq.~\ref{zerocross} with z as the fit parameter.  We find a shift of $\Delta z = 0.19$ which is well within the average FWHM $\approx 2.35$ found above.  Using the best fit $z$ we find that the shape of the boundary is fairly accurate with RMS $\approx 0.15$ which is again well within the average FWHM.  The error is largely due to the width of the $\braket{Z_0}(\omega)$ dips and the influence of nearby transitions.  As seen in Fig.~\ref{Fnew} the [0,1] dip is pulled to higher $\omega$ by the [2,3] transition near the phase boundary while it is pulled down by the [1,0] transition after the phase boundary.  This causes the boundary to appear to be shifted to higher values of $y$ (or $z$ if $y$ is taken as fixed).  There may also be an effect due to an unintended $ZZ$ rotation, which is a well-known error for quantum computers built from fixed-frequency transmons~\cite{Kandala2020, Sundaresan2020}. While it is difficult to disentangle all of the sources of error, our goal is not to remove the error completely but to demonstrate the procedure for finding the phase boundary using our spectroscopic eigensolver. The accuracy is determined by the width of the dips, set by the time $t$ for which the algorithm can run, which is limited by the number of quantum gates which can be applied before decoherence becomes large.  As quantum hardware improves, this number of gates will increase and the accuracy of our algorithm will improve.

While the energy transitions are symmetric between positive and negative values of $m$, it is often the case that a particular transition is more apparent in the data for either $\pm$ values of $m$.  Fig.~\ref{phase-boundary} is generated by taking the most accurate data between positive an negative $m$. For the raw data and a detailed discussion of how the data was filtered see Supplement~V.

The experiments discussed here demonstrate the ability of the novel spectroscopic eigensolver technique to solve for eigenspectra on near term quantum devices. Similar to the experimental method of tunneling spectroscopy, the simulation register interacts with a local probe qubit to determine its spectrum. This requires only the realization of two-qubit interactions between the probe and register without the coherent control of the register at the heart of many other simulation algorithms, making this technique highly amenable to exploration on near-term noisy quantum computing hardware.

The authors acknowledge the use of IBM Quantum Services for this work. The authors also thank C.S. Hellberg, L.D. Gunlycke, D.J. Egger, S.M. Frolov, P. Jurcevic, and Y. Kim for insightful discussions, D.T. McClure, P. Nation, S. Panda and R. Woo for assistance in performing the experiments, and K.J. Ferris, O.T. Lanes, and K.J. Sung for careful readings of the manuscript.  We acknowledge the Air Force Research Laboratory (AFRL) for providing additional quantum resources through its partnership with IBM.

\bibliographystyle{apsrev4-1}
\bibliography{mzm-phase-refs}

\pagebreak

\section{Hamiltonian Simulation Algorithms}

\begin{figure}[t]
\begin{flushleft}
(a)
\end{flushleft}
\[
\Qcircuit @C=1em @R=0.8em {
  \lstick{\ket{0}} & \gate{H} & \ctrl{3} & \qw & \cdots & & \qw & \multigate{2}{QFT^\dagger} & \meter \\
   & \cdots & & & & & \cdots & \nghost{QFT^\dagger} & \cdots  \\
  \lstick{\ket{0}} & \gate{H} & \qw & \qw & \cdots & & \ctrl{1} & \ghost{QFT^\dagger} & \meter \\
  \lstick{\ket{\psi_0}} & {/} \qw & \gate{U_{\rm Pauli}^{N_t 2^{m-1}}}  & \qw & \cdots & & \gate{U_{\rm Pauli}^{N_t 2^0}} & \qw & \qwa
}
\]

\begin{flushleft}
(b)
\end{flushleft}
\[
\Qcircuit @C=1em @R=0.8em {
  \lstick{\ket{0}} & \qw & \qw & \gate{R_0^z(-\omega\,dt)} & \qw & 
    \multigate{1}{R_{0i}^{xx}(2c\,dt)} & \qw & \qw &  \meter \\
 \lstick{} & \qw & \qw & \multigate{2}{U_{\rm Pauli}} & \qw &
    \ghost{R_{0i}^{xx}(2c\,dt)} & \qw & \qw & \qwa \\
    & & \cdots & \nghost{U_{\rm Pauli}} & \cdots & & & & \cdots \\
  \lstick{} & \qw & \qw & \ghost{U_{\rm Pauli}} & \qw & \qw & \qw & \qw & \qwa
    \inputgroupv{2}{4}{0.7em}{1.69em}{\ket{\psi_0}}
    \gategroup{1}{3}{4}{6}{2.4em}{(}
    \gategroup{1}{3}{4}{6}{2.4em}{)}
}
\raisebox{3.4ex}{\makebox[-21pt][r]{$N_t$}}
\]
\caption{Comparison of (a) QPE and (b) the spectroscipic eigensolver algorithms \label{qpe-vs-sea}.}
\end{figure}
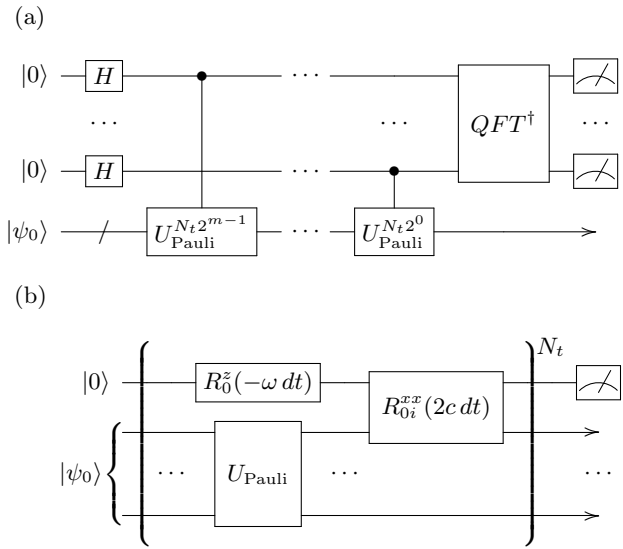

Quantum computers will theoretically give an exponential speed-up to the problem of solving for the eigenvalues of Hamiltonians by the celebrated quantum phase estimation (QPE) algorithm~\cite{Kitaev1995} -- the backbone of proposed universal quantum simulation~\cite{Lloyd1996}. QPE estimates the phase of a unitary operator $U$ acting on an $n$-dimensional Hilbert space encoded by a ``simulation'' register to within an arbitrary precision specified by the number $m$ of qubits in an auxiliary register. Performing QPE involves realizing controlled-$U^k$ gates (Trotterized into $N_t$ steps each) on $n$ qubits (with $1 \le k \le 2^{m-1}$) followed by an inverse quantum Fourier transform on the auxiliary register (as depicted in Fig.~\ref{qpe-vs-sea}a), and is prohibitive for near-term noisy quantum hardware~\cite{Preskill2018}. Current research thrusts involve reducing the quantum resources necessary for performing quantum simulations while still seeking exponential speed-up. One such method known as iterative phase estimateion (IPE) replaces the $m$-qubit auxiliary register with a recycled ``pointer'' qubit via mid-circuit measurements and feed-forward~\cite{Corcoles2021}. Another method involves performing a classical Fourier transform on the time series generated by expectation values of the phase kickback on the auxiliary qubit from a controlled-$U$ on the simulation register~\cite{Somma2019}. The spectroscopic eigensolver (Fig.~\ref{qpe-vs-sea}b for comparison), however, uses only a local interaction and does not require coherent control of the register to elucidate the eigenvalues. As a demonstration of the expected performance, we will estimate a fidelity score for the resource extremal case of QPE.

The accuracy of the spectroscopic eigensolver on the Landau-Zener model is approximately equivalent to three bits of precision ($\sim 10\%$) and is obtained with a single qubit simulation register.  The same precision in QPE requires a three-qubit measurement register. Using the \texttt{PhaseEstimation} circuit from Qiskit with the same Trotterization, SWAP-mapping determined by the (stochastic) SABRE SWAP method of the Qiskit transpiler, the resulting circuit is mapped to the least noisy qubits on \textit{ibm\_lagos} with the \texttt{mapomatic} package~\cite{mapomatic}. The \texttt{mapomatic} package provides an infidelity score calculated from single- and two-qubit gate error and measurement error, with each $R^{zx}$ gate error calculated as linear in angle (as found in Ref.~\cite{Stenger2021}) with a floor of the $X$-error that forms the echo. These are reported in Table~\ref{table:qpe-vs-sea}. We see that QPE is estimated to have three-9's of infidelity while the spectroscopic eigensolver is around 65\%, which improves to 6\% when pulse-scaled $R^{zx}$ gates are used. Note that while these exact values are based on current error rates determined by calibration and are subject to frequent change, they provide an estimate to the required quantum resources.

\begin{table}
\centering
\begin{tabular}{l||c|c|c}
 & QPE & Spec. Eigen. & Spec. Eigen. w/ $R^{zx}$ \\ \hline
$N_{\rm qubits}$ & 4 & 2 & 2 \\
\#CNOTs & 1272 & 120 & $60^*$ \\
Infid. Score & 0.99984 & 0.65363 & 0.06018 \\
\end{tabular}
\caption{A comparison of parameters between QPE and the spectroscopic eigensolver for the Landau-Zener model, and that implemented with $R^{zx}(\theta)$ operations, the count of which is denoted in the table by an asterisk (*). \label{table:qpe-vs-sea}}
\end{table}

\section{Functional form of the resonance eigensolver}
\label{FFRE}

The algorithm works by applying unitary time evolution to an arbitrary state (in our simulation we choose $|\psi_0 \rangle = |00\rangle$) of the system with the probe qubit in the ground state.  We then observe the probability that the probe qubit becomes excited.  If the energy on probe qubit (as swept by $\omega$ in the resonance Hamiltonian, not the actual transition energy of transmon as in~\cite{Roushan2017}) matches an energy transition in the system then we will find a peak in the probability. To demonstrate that the probability indeed peaks at energy transitions we will use two approaches. First we will derive the first order perturbation of the probability as a function of probe qubit frequency.  This will show us that the probability decays as the frequency of the probe qubit is taken off resonance.  Second we will analyze the behavior of the probability when the system is composed of only two energy levels.  Since the probability decays away from resonance, a two level system is a good approximation for a larger system where a particular transition is close to resonance with the probe qubit.  

Let us start with the perturbation approach.  In the algorithm, we apply unitary time evolution
\begin{equation}
    U_{\rm res}(\omega) = e^{-i (- \omega Z_0/2 + H_{\rm Pauli} + c X_0 X_i) dt}
\end{equation}
where $H_{\rm Pauli}$ is the qubit-encoded system Hamiltonian, qubit zero is the probe qubit, and qubit $i$ is the qubit being probed.  We desire $c$ to be small so that we can treat the $c X_0 X_i$ term as a perturbation.  Let $E_{n}$ and $\ket{n}$ be the energies and eigenvectors of the system  Hamiltonian $H_{\rm Pauli}\ket{n} = E_{n}\ket{n}$ so that the unperturbed eigenvalues are $E_{a,n} = (-1)^{a+1} \omega/2 + E_{n} $  and the unperturbed eigenvectors are $\ket{a,n}$ where $a\in\{0,1\}$ labels the state of the probe qubit, i.e.
\begin{equation}
\begin{split}
    \left(-\frac{1}{2} \omega Z_0 + H_{\rm Pauli}\right) \ket{a,n} &= E_{a,n}\ket{a,n} 
    \\
    &= \left(\frac{(-1)^{a+1}}{2}\omega+E_n\right)\ket{a,n}.
\end{split}
\end{equation}
Let $\bar{E}_{a n}$ be the energy of the entire resonance Hamiltonian and $\ket{\overline{an}}$ be the eigenvectors so that
\begin{equation}
\begin{split}
    \left(- \frac{1}{2} \omega Z_0 + H_{\rm Pauli} + c X_0 X_1\right) \ket{\overline{an}} = \bar{E}_{an}\ket{\overline{an}}
\end{split}
\end{equation}
To first order in $c$ we have,
\begin{equation}
    \begin{split}
        &\bar{E}_{a n} \approx E_{a n} + c \bra{a,n}X_0X_1\ket{a,n} = E_{a n}
        \\
        &\ket{\overline{an}} \approx \ket{an} + c \sum_{bm\neq an} \ket{bm}\frac{\bra{bm}X_0X_1\ket{an}}{E_{a n} - E_{b m}}.
    \end{split}
\end{equation}

The initial state of the algorithm has the probe qubit in the ground state and the system qubits in an arbitrary state.  We are interested in the probability that the auxiliary qubit will flip regardless of the state of the system after time evolution.  We can write the entire algorithm in one line,
\begin{equation}
    \braket{Z_0}(\omega) = 1 - 2\sum_m\left|\sum_{n}\alpha_{n}\bra{1m}U_{\rm res}(\omega)^{N_t} \ket{0n}\right|^2
\end{equation}
where the parameters $\alpha_n$ are arbitrary reflecting that we start in an arbitrary state of the system.  To evaluate this probability, we need to evaluate 
\begin{equation}
    \begin{split}
        \bra{1m}\left(-\frac{1}{2} \omega Z_0 + H_S + c X_0 X_1\right)^l\ket{0n} =
        \\
        \sum_{ak} \braket{1m | \overline{ak}}\braket{\overline{ak}|0n}\bar{E}_{ak}^l
    \end{split}
\end{equation}
where $l$ is any non-negative integer. To first order, we have
\begin{equation}
    \begin{split}
        \braket{1m | \overline{ak}} = \delta_{a1}\delta_{km} + c   \frac{\chi_{mk}}{ E_{0m} - E_{1k}}\delta_{a0}
    \end{split}
\end{equation}
\begin{equation}
    \begin{split}
        \braket{\overline{ak} | 0n} = \delta_{a0}\delta_{kn} + c   \frac{\chi_{kn}}{ E_{1k} - E_{0n}}\delta_{a1}
    \end{split}
\end{equation}
where $\chi_{mn} = \bra{m}X_1\ket{n}$.
We multiply these together and get,
\begin{equation}
    \begin{split}
        \bra{1m}\left(-\frac{1}{2}\omega Z_0 + H_S + c X_0 X_1 \right)^l\ket{0n} =
        \\
        c \chi_{mn} \left(   \frac{E_{1m}^l}{ E_{1m} - E_{0n}} +    \frac{E_{0n}^l}{ E_{0m} - E_{1n}}  \right).
    \end{split}
\end{equation}
Now we can go back to the probability
\begin{equation}
\begin{split}
    &\braket{Z_0}(\omega) = 
    \\
    &1-2 c^2 \left|\sum_{mn}\alpha_n\chi_{mn}\left(   \frac{e^{iE_{1m}t}}{ E_{m} - E_{n}+\omega} +    \frac{e^{iE_{0n}t}}{ E_{m} - E_{n}-\omega}  \right) \right|^2    
\end{split}
\end{equation}
We see that the biggest contributions to the probability are the energies which are on resonance with the probe qubit.  The pole at  $\omega = \pm (E_m -E_n)$ is unphysical since the perturbation is only valid for $c < E_m - E_n - \omega$ for all $m$ and $n$. 

To analyze the behavior near the resonance condition $\omega = \pm(E_1 -E_0)$, we can ignore the other energy levels since they do not contribute as strongly.  In this case we are left with a two level system.  We can fully analytically analyze our algorithm for this two level system which will tell us generally the behaviour of the algorithm when $\omega$ is near an energy transition.  Let us write the two level system Hamiltonian as,
\begin{equation}
    H_0 = \frac{1}{2} d Z_1
\end{equation}
where $d = E_1 - E_0$.  The full Hamiltonian is,
\begin{equation}
    H = \frac{1}{2} \omega Z_0 + \frac{1}{2} d Z_1 + c X_0 X_1
\end{equation}
This has eigenvalues
\begin{equation}
\begin{split}
        E_{a}^{\pm} = \pm \frac{1}{2} \sqrt{4c^2 + (d-\omega)^2}
\\
E_{b}^{\pm} = \pm \frac{1}{2} \sqrt{4c^2 + (d+\omega)^2}
\end{split}
\end{equation}
and eigenvectors
\begin{equation}
    \begin{split}
        &|a \pm \rangle =  \left( \frac{d-\omega + 2E_a^{\pm}}{2N_a^{\pm}}|01 \rangle - \frac{c}{N_a^{\pm}} |10 \rangle  \right)
\\
        &|b \pm \rangle  =  \left( \frac{d+\omega + 2E_b^{\pm}}{2N_b^{\pm}}|00 \rangle + \frac{c}{N_b^{\pm}} |11 \rangle \right)
    \end{split}
\end{equation}
where
\begin{equation}
    \begin{split}
        &N_{a}^{\pm} = \frac{1}{\sqrt{2}}\sqrt{(d-\omega)^2+4c^2-2(d-\omega)E_{a}^{\pm}}
\\
        &N_{b}^{\pm} = \frac{1}{\sqrt{2}} \sqrt{(d+\omega)^2+4c^2+2(d+\omega)E_{b}^{\pm}}
    \end{split}
\end{equation}
Thus, we can write time evolution as,
\begin{equation}
\begin{split}
        e^{-iHt} &=  e^{-iE_a^-} |a-\rangle\langle a-| + e^{-iE_a^+} |a+\rangle\langle a+| 
        \\
        &+ e^{-iE_b^-} |b-\rangle\langle b-| + e^{-iE_b^+} |b+\rangle\langle b+|
\end{split}
\end{equation}
Using this form of the time evolution operator it is straight forward (although tedious) to derive the bracket,
\begin{equation}
    \langle 1m|e^{iH(\omega)t}|0n \rangle = A(\omega) \delta_{n1}\delta_{m0} + B(\omega) \delta_{n0}\delta_{m1}
\end{equation}
where 
\begin{equation}
\begin{split}
    A(\omega,d)=  i \frac{2c }{\sqrt{4c^2 + (d-\omega)^2} } \sin(E_a t)
    \\
    B(\omega,d) =  i \frac{2c }{\sqrt{4c^2 + (d+\omega)^2} }  \sin(E_b t)
\end{split}
\end{equation}
Once again, we see that the probability is maximized when $\omega = \pm d$.  

\begin{figure}[h]
\begin{center}
\includegraphics[width=\columnwidth]{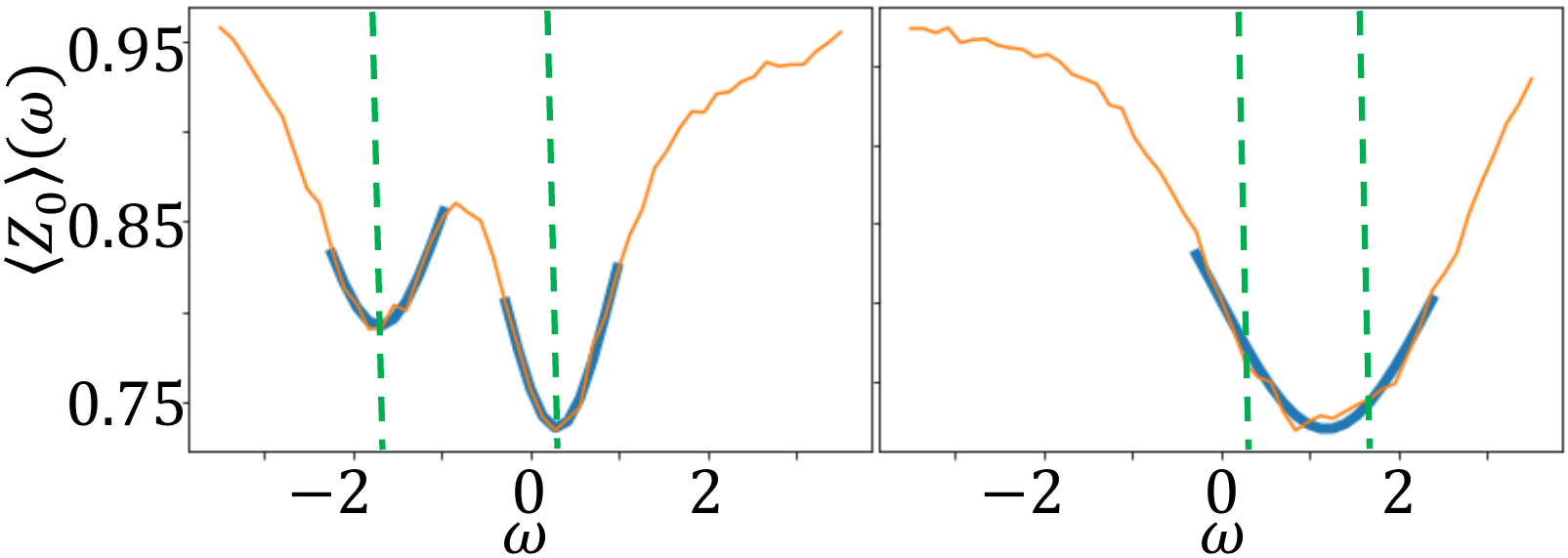}
\end{center}
\vspace{-2mm}

\caption{Fit of the {\it ibm\_lagos} data to the functional form.  The orange curves are the experimental data, the blue curves are the fit, and the green dashed lines show the location of the transitions according to the fit.}
\label{fit}
\vspace{3mm}
\end{figure}

We use the functional form to assist in finding the minima in the data.  Figure~\ref{fit} shows the functional form fitted to the data from {\it ibm\_lagos}. We begin the fitting by smoothing the data (averaging over the nearest four data points) and then finding the zeros in the derivative.  We expect two energy transitions within the data range, specifically the [0,1] transition and the [2,3] transition which are both small. If we find two zeros (left panel in Fig.~\ref{fit})) then we take the ten nearest points to each minimum found by the derivative and separately fit them to $|A(\omega,d)|^2$.  If we find only one minimum (right panel in Fig.~\ref{fit}) then we take the 20 nearest points and fit it to $|A(\omega,d1) + A(\omega,d2)|^2$.  In other words, we expect that the two transitions are close enough that they form a single dip.

\section{Zero energy resonance dip}

In Fig.~2 of the main text, we see a dip in $\braket{Z_0}(\omega)$ at $\omega = 0$.  However, this peak is absent in the Kitaev chain data.  To understand the origin of the zero-peak let us return to the function form of the probe qubit orientation as derived in appendix~\ref{FFRE}
\begin{equation}
\begin{split}
    &\braket{Z_0}(\omega) = 
    \\
    &1-2 c^2 \left|\sum_{mn}\alpha_n\chi_{mn}\left(   \frac{e^{iE_{1m}t}}{ E_{m} - E_{n}+\omega} +    \frac{e^{iE_{0n}t}}{ E_{m} - E_{n}-\omega}  \right) \right|^2 .   
\end{split}
\end{equation}
We see that there is a dip in $\braket{Z_0}(\omega)$  when $\omega = E_m - E_n$.  Since the sum is over all $m$ and $n$ we should expect a zero energy dip for $m=n$ unless $\chi_{nn} = 0$.  Recall that 
\begin{equation}
    \chi_{mn} = \bra{m}X_1\ket{n}
\end{equation}
which may not be zero depending on the eigenvectors $\ket{n}$ of the Hamiltonian.  If, however, the Hamiltonian describes fermions and we are using the Jordan-Wigner transformation as in the main text then $X_1$ creates or destroys a single fermion.  Since fermion Hamiltonians conserve parity, they always have $\chi_{nn} = 0$.  Therefore, we should not expect a zero dip for fermion Hamiltonians.  If, however, there is some small error in term which couples the system to the probe when implemented on the quantum hardware then a zero dip can show up.  For this work we were careful to find qubits that which did show a zero peak for the fermion model.

\section{Optimization}

\begin{figure}[h]
\begin{center}
\includegraphics[width=\columnwidth]{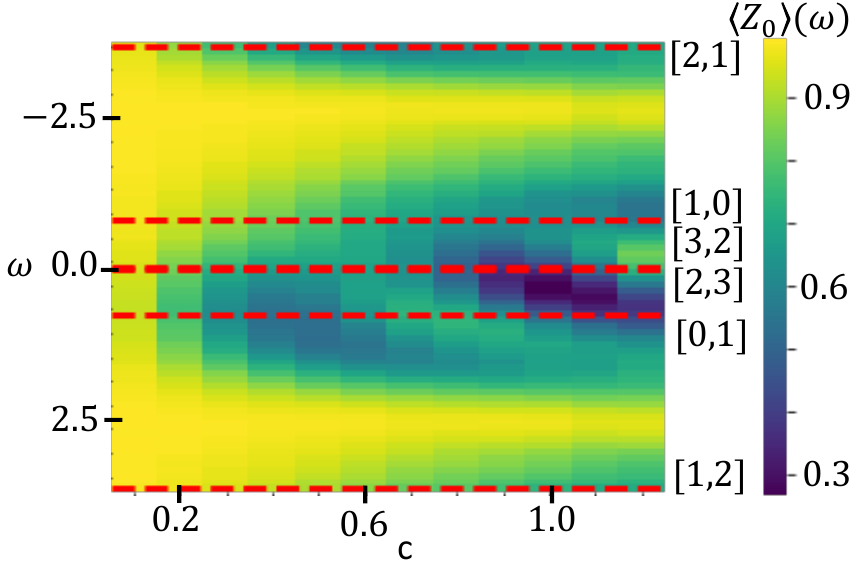}
\end{center}
\vspace{-2mm}

\caption{Expectation values of the $Z$-component of the probe qubit as a function of $\omega$ and $c$.  The other parameters are set to $x=1.5$, $y = 0.4$, $z=-0.3$, $t = 5.0$, $dt = 1.2$.  The dashed red lines show the location of the energy transitions.  The lines are labeled on the right of the figure by the two states involved in the transition, e.g. $[0,1]$ is the transition from state $0$ to state $1$. }
\label{c_op}
\vspace{3mm}
\end{figure}

\begin{figure}[h]
\begin{center}
\includegraphics[width=\columnwidth]{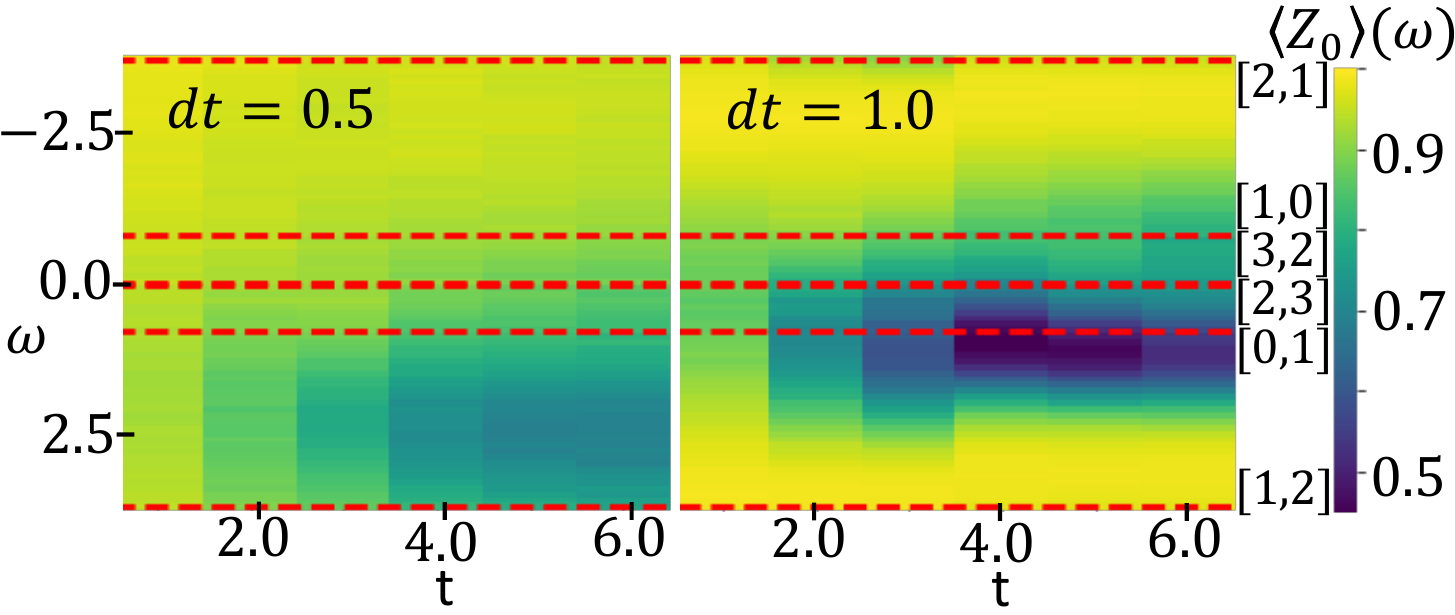}
\end{center}
\vspace{-2mm}

\caption{Expectation values of the $Z$-component of the probe qubit as a function of $\omega$ and $t$ for $dt = 0.5$ (left) and $dt = 1.0$ (right).  The other parameters are set to $x=1.5$, $y = 0.4$, $z=-0.3$, $t = 0.4$.  The dashed red lines show the location of the energy transitions.  The lines are labeled on the right of the figure by the two states involved in the transition, e.g. $[0,1]$ is the transition from state $0$ to state $1$.   }
\label{t_op}
\vspace{3mm}
\end{figure}

The coupling strength $c$ between the probe qubit and the system, the total evolution time $t$, and the time step $dt$ all need to be optimized for the algorithm.  In Fig.~\ref{c_op} we see the orientation of the probe qubit as a function of both $\omega$ and $c$.  We want to focus on the ground to first excited state transition labeled $[0,1]$ in the figure.  We see that when $c>0.5$ the dip corresponding to the $[0,1]$ transition is shifted due to the large interaction between the probe and the system, and in fact the reverse transition $[1,0]$ and higher-order transitions $[2,3]$ and $[3,2]$ also appear (yet also shifted) at these strong coupling values.  However, for $c < 0.1$ the dip is washed out. We choose $c=0.3$ where the dip corresponding to the correct transition ($[0,1]$) is apparent and at the correct value, and no other nearby transitions are observed.  

Once the optimum $c$ value was found we optimized time step $dt$ and total time $t$ in a similar way.  Then we re-optimized $c$ at the new values of $t$ and $dt$ and found that the optimum $c$ value had not changed.  Figure~\ref{t_op} shows the $Z$-expectation of the probe qubit as a function of $\omega$ and $t$ for two values of $dt$.  We see that if $t$ is too short the dip is washed out just like for small $c$.  Additionally, if $dt$ is too small then we run into errors due to the increased number of gates that are applied to the quantum register. We want the smallest $dt$ we can manage so that our Suzuki-Trotter decomposition is as accurate as possible.  

Informed by these optimization experiments and others, we chose the parameter set $t=5.0$, $dt = 0.7$, $c = 0.3$ for this work. More data was analyzed than those presented here, and we have made our full data set accessible at \url{www.github.com/qiskit-research/mzm-phase-boundary}.

\section{Applying the chemical potential symmetry}

\begin{figure}[h]
\begin{center}
\vspace{2mm}
\includegraphics[width=\columnwidth]{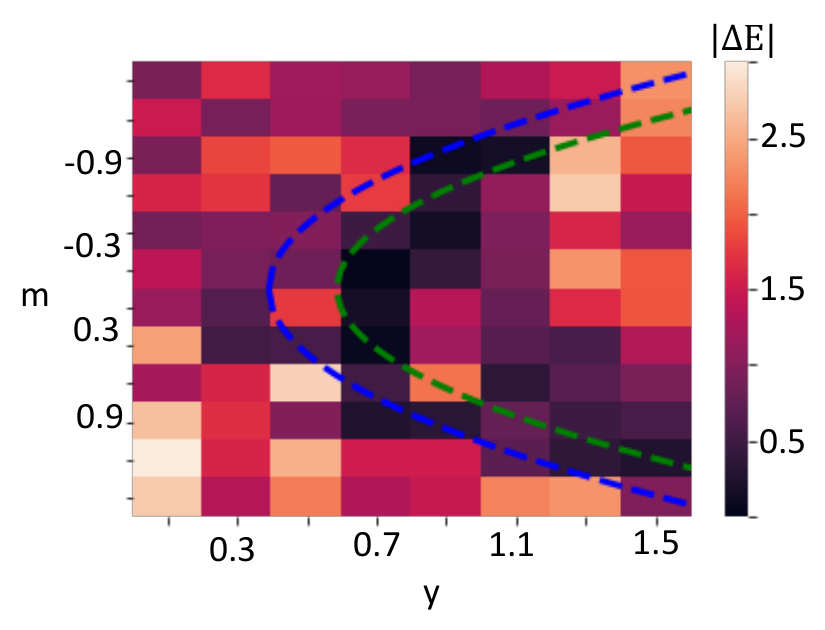}
\end{center}
\vspace{-2mm}

\caption{The raw data of the energy gap as a function of $m$ and $y$.  The parameter values match those in Fig.~\ref{phase-boundary} of the main text.  The dashed blue curve shows the location of the phase transition calculated from Eq.~\eqref{zerocross} of the main text while the green curve is a best fit of the data to Eq.~\eqref{zerocross} using $z$ as the fit parameter.  }
\label{m-symm}
\vspace{3mm}
\end{figure}

We know that the energy spectrum is symmetric about $m\rightarrow-m$.  Therefore, if we know the zero-crossing phase boundary for $+m$ we can infer the phase boundary for $-m$ and visa-versa.  Figure~\ref{m-symm} shows the raw data for the energy gap as a function of $m$ and $y$.  In the main text we selected the smallest gap between positive and negative $m$ to be the gap for both.  In this way, we filter out the data in which the [0,1] transition has been washed out by the [2,3] transition.    Sometimes the $[2,3]$ transition has a much larger response than the $[0,1]$ transition.  When this happens the $[2,3]$ transition can wash out the $[0,1]$ transition so that the fitting technique fails to provide an accurate estimate of the $[0,1]$ transition.  Thus, we get more accurate data by selecting between the transitions at positive and negative $m$.  We find that the zero crossing of the $[0,1]$ transition is easily extractable from the data for negative $m$ in the range of $0.7 \leq y \leq 1.1$ while for $1.1 < y$ the positive $m$ values show the $[0,1]$ zero-crossing clearly. 

\section{Approaching the $n\rightarrow\infty$ limit in the Kitaev chain}

\begin{figure}[t]
\begin{center}
\includegraphics[width=\columnwidth]{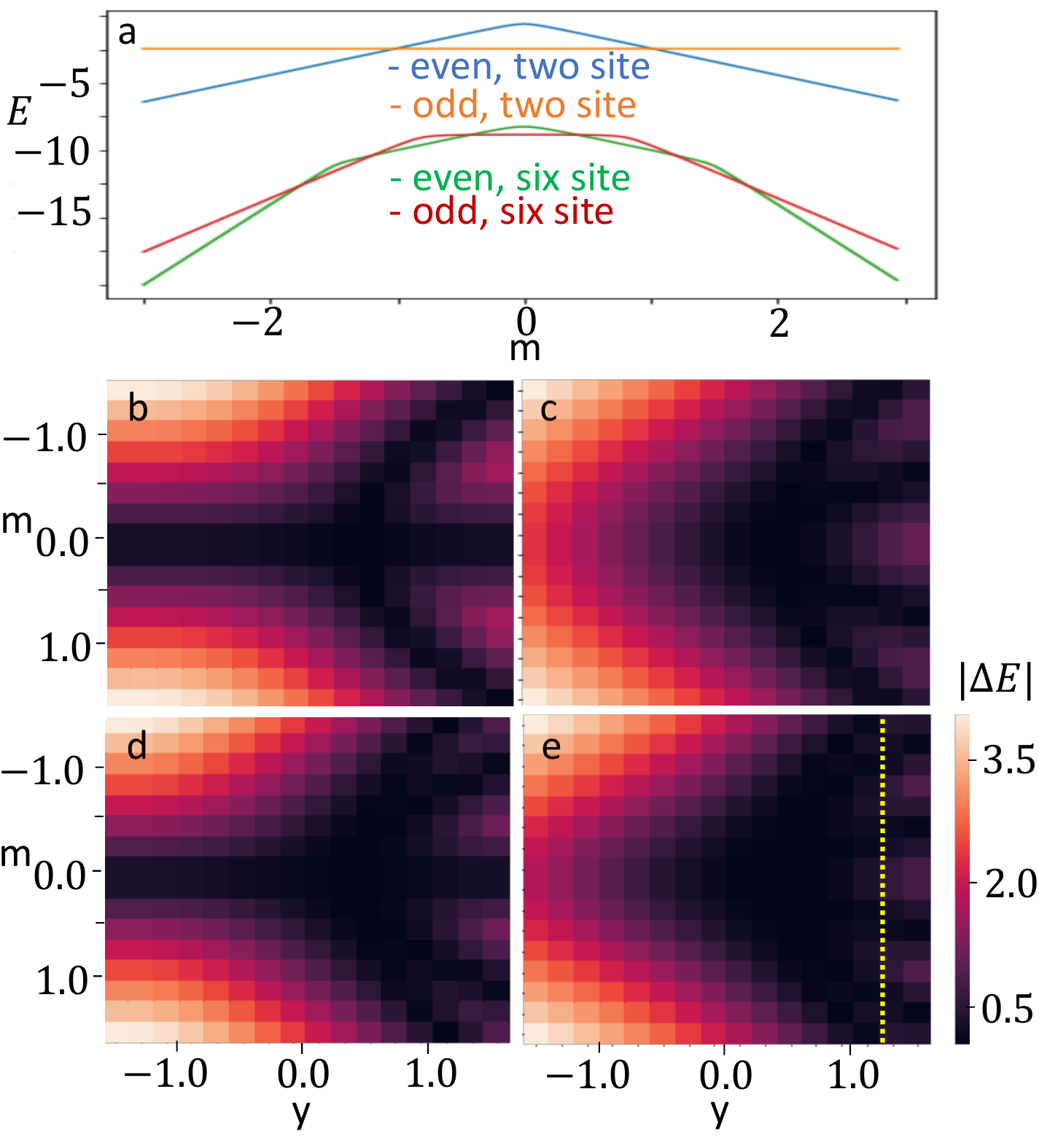}
\end{center}
\vspace{-2mm}

\caption{Energy gap for increasing chain length.  The plots are generated from a classical simulation.  In panel (a) we plot the (absolute) energy of the lowest even and lowest odd parity states versus $m$ for two sites and for six sites at $y=1.3$ ($m$ and $y$ are defined in Eq.~\eqref{zerocross}).  In the other panels, the energy gap between the lowest even and lowest odd parity state is plotted against the parameters $m$ and $y$.  The points where $|\Delta E|$ goes to zero corresponds to the crossing of the two lowest energy levels.  Each plot has a different number of sites: (b) three sites, (c) four sites, (d) five sites, (e) six sites, the dotted yellow line shows the location ($y=1.3$) of the six site energy cuts in panel (a). For a plot of the simulated 2-site model, see Fig.~\ref{phase-boundary}a.  }
\label{vsL}
\vspace{3mm}
\end{figure}

We can approach the $L\rightarrow \infty$ limit by increasing the number of sites.  In Fig.~\ref{vsL} we plot maps of the energy gap for different values of $L$.  The parameters are set to $x = 1.5$, $z=\bar{m} = 0.4$. For $L>2$, we see that the gap opens and closes multiple times.  The larger $L$ the more the gap closes and the smaller are the peaks between the lines where the gap closes.  In this way, one could approach the case where the gap stays closed.

\section{Two site party flipping boundary}

In the main text we stated that the parity of the ground state changes along the boundary
\begin{equation}
    m = \sqrt{z^2 + z(x+y)+ xy}
\end{equation}
for the two site Kitaev chain.  This comes simply from diagonalizing the two site Hamiltonian
\begin{equation}
    H = m Z_1 + m Z_2 + x X_1 X_2 + y Y_1 Y_2 + z Z_1 Z_2
\end{equation}
which has eigenvalues
\begin{equation}
\begin{split}
    &E_a^{\pm} = -z \pm (x+y) \\
    &E_b^{\pm} = z \pm \sqrt{4m^2 + (x-y)^2}
\end{split}
\end{equation}
The parameter regime of interest, $E_a^-$ and $E_b^-$ are the two lowest energy levels.  We want to know when the ground state switches from $E_a^-$ to $E_b^-$.  Setting $E_a^- = E_b^-$ and solving for $m$ gives the desired equation.

\section{Three site model data}

\begin{figure}[h]
\begin{center}
\vspace{2mm}
\includegraphics[width=\columnwidth]{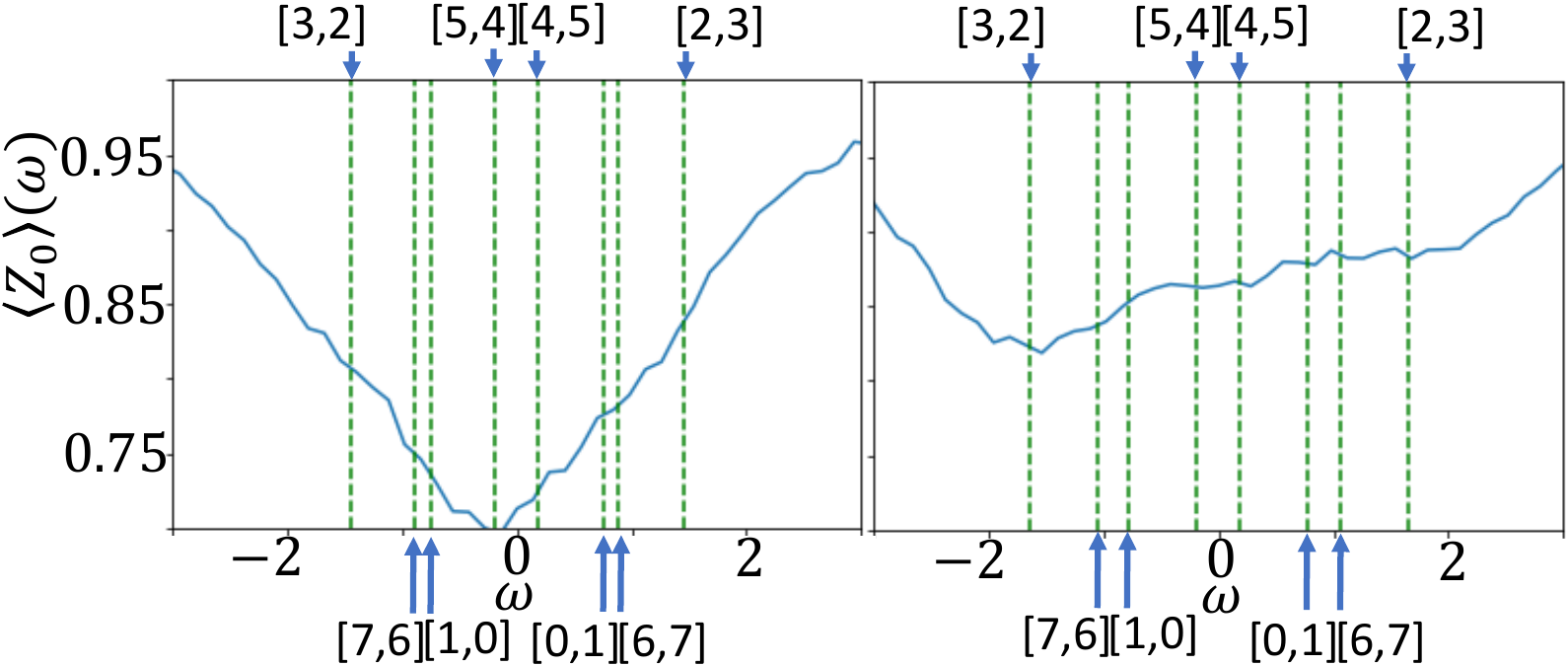}
\end{center}
\vspace{-2mm}

\caption{Expectation of the $Z$-component of the probe qubit as a function of $\omega$ for a three-site Kitaev chain.  Left: for $m = 0.5$.  Right: for $m = -0.5$.  The other parameters are set to $x = 1.5$, $y = 1.1$, $z = 0.4$, $c = 0.3$, $dt = 0.7$, $t = 5$.  The green dashed lines show the energy transitions which are labeled at the top/bottom of the plot.}
\label{3site1}
\vspace{3mm}
\end{figure}

\begin{figure}[b]
\begin{center}
\includegraphics[width=\columnwidth]{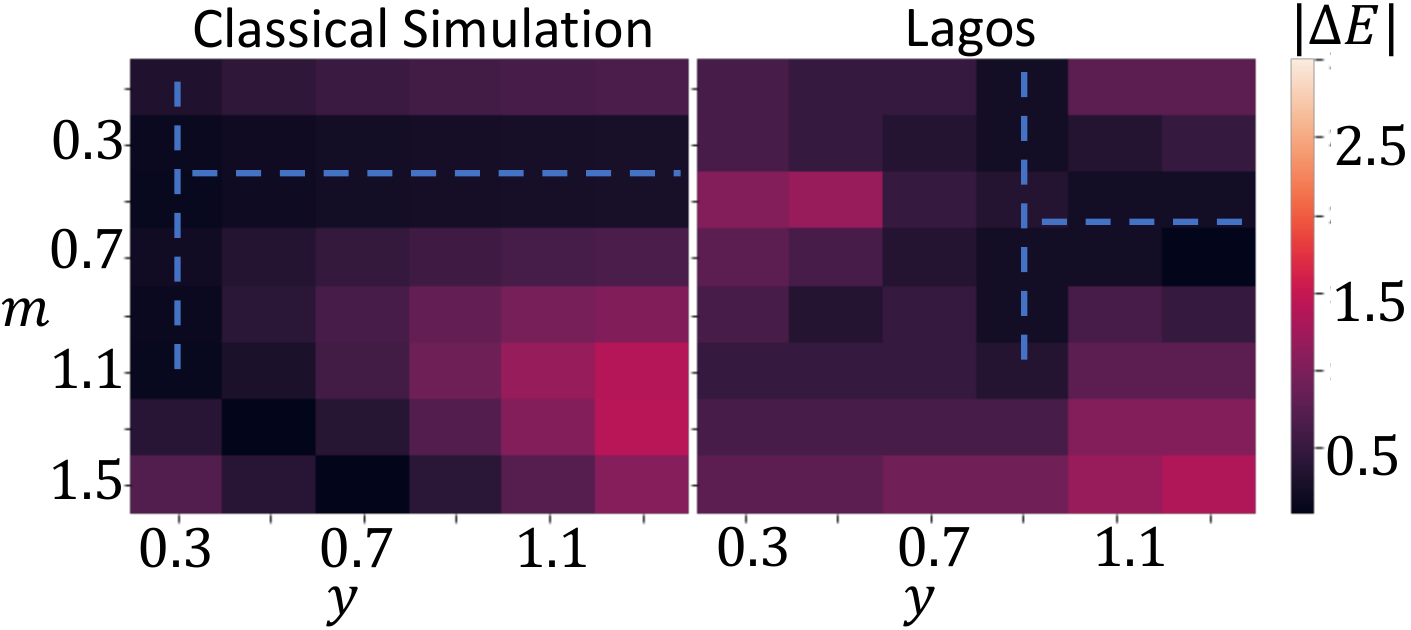}
\end{center}
\vspace{-2mm}

\caption{ The $[5,4]$ energy transition as a function of $y$ and $m$.  Left: from directly solving for the eigenvalues of the Hamiltonian.  Right: from the spectroscopic algorithm ran on Lagos.  The other paramaters are set to $x = 1.5$, $z = 0.4$, $c = 0.3$, $dt = 0.7$, $t = 5$. }
\label{3site2}
\vspace{3mm}
\end{figure}

We ran the spectroscopic algorithm for a three site Kitaev chain.  While the data is very accurate at certain parameter values, we where not able to map out the full phase diagram.  One issue is that the number of small-energy transitions increases causing the width of the dips in $\braket{Z_0}(\omega)$ to often absorb several of the weaker transitions.  In Fig.~\ref{3site1} we plot $\braket{Z_0}(\omega)$ for $m = 0.5$ (left) and $m = -0.5$ (right).  While the energy transitions are symmetric in $m$, the algorithm often favors different transitions for positive and negative $m$.  In this case, the $[5,4]$ transition is favored for positive $m$ while $[3,2]$ is favored for negative $m$.  While this asymmetry in the favored transition is present for the two site case as well, there we where able to resolve the secondary transition for many parameter values.  For the three site model, the secondary transitions are very difficult to resolve if they are present at all.

Still, we are able to trace out the zero crossings for certain energy transitions for a range of parameters.  Figure~\ref{3site2}
shows the zero-crossing for the $[5,4]$ transition in a small range of $m$ and $y$.  While the transition is shifted just like in the two site case, the main features are present.  The data could be improved by narrowing the dips in $\braket{Z_0}(\omega)$.  A possible solution for narrowing the dips would be to increase $t$, which for this current work was limited by qubit coherence and classical waveform generation bandwidth for {\it ibm\_lagos}. It is expected that future generations of superconducting quantum backends will allow for such exploration. 

\end{document}